\documentclass[aps,prb,a4paper,reprint,superscriptaddress,showpacs]{revtex4-1}
\usepackage{graphicx,graphics,color,epsfig}
\usepackage{epstopdf} %This line makes .eps figures into .pdf - please comment out if not required.
\usepackage{bm}
\usepackage{hyperref}
\usepackage{amsmath}
\usepackage{amsfonts}
\usepackage{amssymb}
\usepackage{appendix}
\usepackage{booktabs}

\begin{document}

\title{Spin-selectable, {region-tunable} negative differential resistance in
graphene double ferromagnetic barriers}

\author{Yu Song}\thanks{Corresponding author}
%\email{songyu@mtrc.ac.cn}
\email{kwungyusung@gmail.com}
\affiliation{Microsystem and Terahertz Research Center, China Academy of Engineering Physics,
Chengdu 610200, P.R. China}
\affiliation{Institute of Electronic Engineering, China Academy of Engineering Physics,
Mianyang 621999, P.R. China}

\author{Yang Liu}
\affiliation{Microsystem and Terahertz Research Center, China Academy of Engineering Physics,
Chengdu 610200, P.R. China}
\affiliation{Institute of Electronic Engineering, China Academy of Engineering Physics,
Mianyang 621999, P.R. China}

\author{Xiaolong Feng}
\affiliation{Microsystem and Terahertz Research Center, China Academy of Engineering Physics,
Chengdu 610200, P.R. China}
\affiliation{Institute of Electronic Engineering, China Academy of Engineering Physics,
Mianyang 621999, P.R. China}

\author{Fei Yan}
\affiliation{Institute of Electronic Engineering, China Academy of Engineering Physics,
Mianyang 621999, P.R. China}

\author{Wei-Zhi Zhang}
\affiliation{Institute of Electronic Engineering, China Academy of Engineering Physics,
Mianyang 621999, P.R. China}

\begin{abstract}
We propose a graphene device that can generate spin-dependent negative differential resistance (NDR). The device is composed of a sufficiently wide and short graphene and two gated EuO strips deposited on top of it. This scheme avoids graphene edge tailors required by previous proposals. More importantly, we find clear significant of a spin selectivity and a {region tunability} in the spin-dependent NDR: by changing the top gates of the device, NDR for spin up only, spin down only, or both spins (occurring sequentially) can be respectively realized; meanwhile, the central position of the NDR {region} in each case can be monotonously tuned in a wide range of bias voltage. These remarkable features are attributed to a gate controllability of the spin-dependent resonant levels in the device hence their deviations from the Fermi energy and Dirac point in the source electrode, respectively. They add a spin and a bias degree of freedom to {conventional} NDR-based devices, which paves a way for designing a whole new class of NDR circuits.
\end{abstract}

%\pacs{ }
\date{\today} %15 February 2017
\maketitle

\section{Introduction}

The intersection of spintronics and nonlinear transport % at finite bias
can lead to phenomena of interest and use.
For example, the realization of a spin-dependent negative differential resistance (NDR),
in which {fermions of one spin display NDR} while the current of the other spin increases monotonically as the bias increases,
can be applied as spin-resolved {oscillators,\cite{alekseev2000large} amplifiers,\cite{laskar1989gate}
switchings,\cite{liu1997novel} and memories}.\cite{van1999tunneling}
%for one of the spins. %\cite{mizuta2006physics}.
Till now, several schemes have been proposed based on graphene nanoribbons.
For example, spin-dependent {NDRs} have been demonstrated in
edge-doped zigzag graphene nanoribbons (ZGNRs),\cite{wu2012negative,jiang2014spin,yan2014effects}
B(N)-doped ZGNRs,\cite{yang2015half}
ZGNRs with nitrogen-vacancy defects,\cite{xu2014negative}
FeN$_4$-embedded ZGNRs,\cite{li2017strong}
and zigzag-edged trigonal graphenes linked to
ZGNRs electrodes.\cite{hong2015axisymmetric} %via carbon atomic chains (CACs).
In these devices, spin-polarized edge states\cite{son2006half} play an important role.
%%%
Other proposals considered armchair graphene nanoribbons (AGNRs),
including {a} FeN$_4$ embedded and N-doped AGNR\cite{liu2016efficient}
and a compound system comprising an AGNR and a set of
ferromagnetic insulator strips deposited on top of it.\cite{munarriz2013strong,diaz2014spin}
%%%
{Very recently, experiments on nanoribbons
%, nanomeshs, nanostructures, and devices
for graphene and other two-dimensional materials have been reported.\cite{bai2010very,Llinas2017short,Chhowalla2016two}
However, proper introductions of magnetic doping's, defects, or superlattices are still required for these devices.
That's partly why till now none of the proposals has been experimentally implemented.
}
%However, till now none of the proposals has been experimentally implemented.
%This is partly because of the extreme difficulty in their fabrications,
%including careful tailors of the graphene edges \cite{bai2010very,Chhowalla2016two,Llinas2017short}and proper introduction of doping's, defects, or superlattices.
%%%%

%

\begin{figure}[!t]
  \centering
  \includegraphics[width=0.75\linewidth]{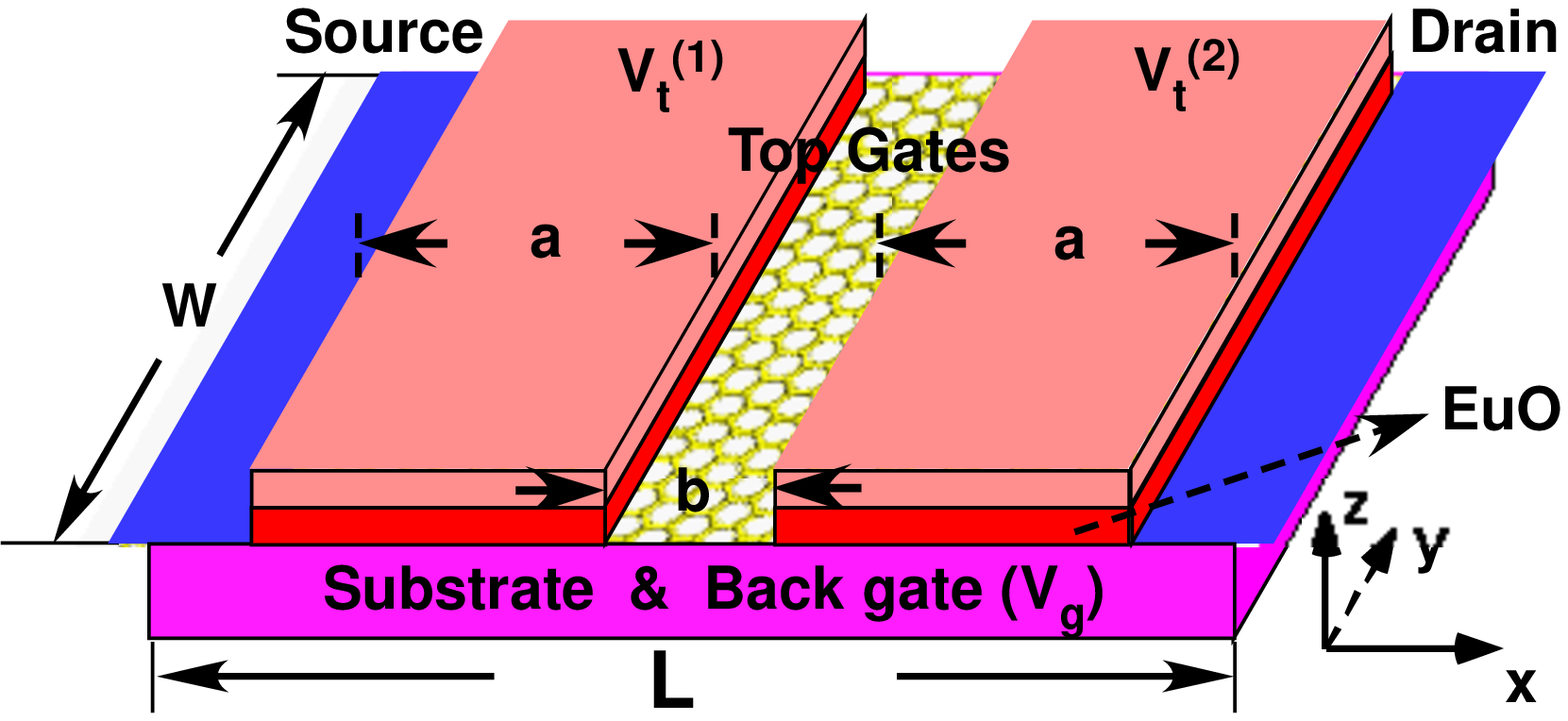}\\
    \includegraphics[width=0.65\linewidth]{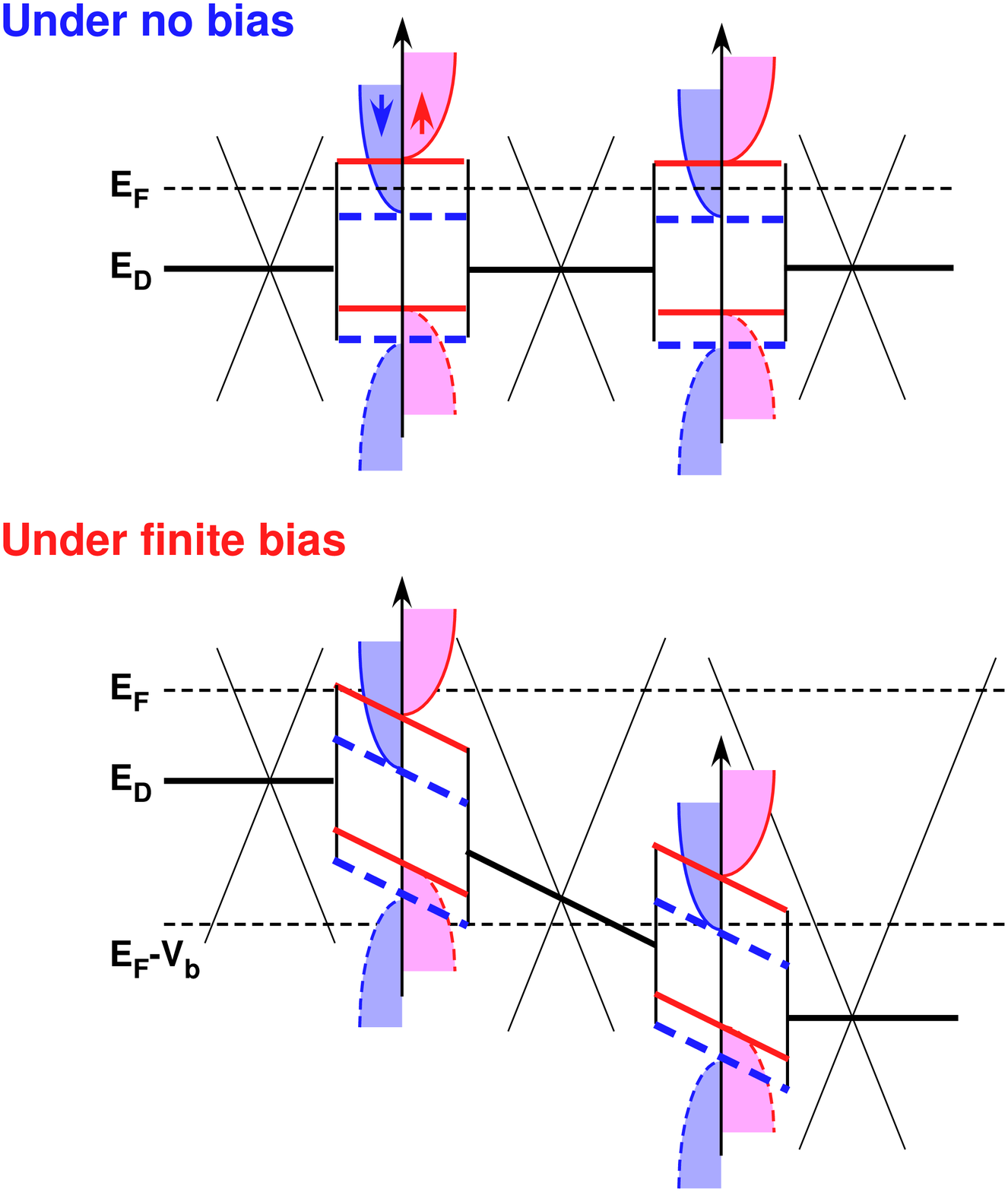}
  \caption{(color online) Schematic diagrams of the proposed device (upper panel)
  and its spin-dependent voltage profiles
  under zero or finite biases (middle or bottom panel).
  Red for spin up and blue for spin down.
  The (spin-dependent) dispersion in each region is also shown.}\label{fig:setup}
\end{figure}

In this work, we propose a bulk graphene based, spin-dependent
double-barrier resonant tunneling diode (DB-RTD),
which requires only
depositing two \emph{gated} EuO strips\cite{haugen2008spin}
on top of a sufficiently wide and short graphene (see upper panel in Fig. \ref{fig:setup}).
{In this device, no specific control of the graphene edge type is needed
because
%the graphene strip is suffiently wide and short and
the transport is dominated by bulk states;
no doping's or defects are required because the ferromagnetism
is induced by magnetic insulators;
and two ferromagnetic barriers instead of a ferromagnetic
superlattice is enough because large energy gaps are induced.}
%We find strong spin-dependent NDR when the Fermi energy is setting around
%one of the spin-dependent resonant levels.
%We demonstrate that in these cases the biased transport
%is dominated by the resonant state, whose energy and integration weight in the current
%decreases with an increasing bias.
We find clear spin-dependent resonant states in the device,
which dominate the biased transport when the Fermi energy is setting around {the
resonant levels}.
The resonant levels decrease with an increasing bias,
resulting in a fewer transversal mode number
%smaller integration weight in the current
hence the occurrence of the desired NDR feature.
%clear significant of spin-dependent resonant states in the device, which
%are shown to dominate the biased transport when the Fermi energy is setting around them.
%As the bias increases, the resonant levels shift to lower values, resulting in
%a decrease of the integration weights in the spin-dependent current %of the resonant state
%and a spin-dependent NDR.
Remarkably,
%More importantly,
we find a clear significant of a spin selectivity and a {region tunability} in the spin-dependent NDR:
by changing the top gates hence the spin-dependent
resonant levels in the device, the spin index of the NDR can be tuned as
spin up only, spin down only, or both spins;
meanwhile, the central position of the NDR {region} in each case can be monotonously tuned
in a wide bias range.
Together with the relative ease in fabrication, these remarkable features
make the proposed device an important building block
for future spintronic or NDR circuits.

The structure of the paper is as follows.
In Sec. \ref{sec:II} we present the setup of the device
and the formula to calculate %its current-voltage characteristics.
the stationary envelop function as well as
the transmission coefficient for a given spin, energy, incident angle, and
bias voltage across the sample.
An ab-initio dispersion of graphene on EuO\cite{yang2013proximity,hallal2017tailoring}
and a linear voltage drop through the device are considered.
The resulting current-voltage (I-V) characteristics of
the device, {comprised of} %spin-dependent resonant states and
a spin-dependent NDR with a spin selectivity and a {region} tunability,
%and a spin current filter effect %the influence of electrode contact doping,
are discussed in the subsequent Sec. \ref{sec:III}.
The mechanism for the arising and tunability of the spin-dependent
NDR, as well as possible applications %of its remarkable features
are emphasized.
Finally, Sec. \ref{sec:IV} concludes the paper.

\section{Setup and Formula}\label{sec:II}

The proposed system is composed of %a back-gated
a sufficiently wide ($W$=250 nm) and short rectangular graphene and
two rectangular EuO strips
on top of it (see the upper panel of Fig. \ref{fig:setup}).
The length of {each strip} is $a =$ 21.75 nm and the spacing between
them is $b =$ 8.7 nm.
The source and drain electrodes, which may induce a charge doping ($U$),
are separated {from the EuO strips by two buffer regions of $a/10$ length}.
The total length $l=2.2a+b$ is several times smaller than $W$, hence the
transport is dominated by bulk states\cite{tworzydlo2006sub}
and no specific control of the graphene edge type is needed.
%%%
Experiments\cite{swartz2012integration,wang2015proximity,wei2016strong,zhang2017realization} and ab-initio calculations\cite{yang2013proximity,hallal2017tailoring} have shown that,
through a magnetic proximity effect, %\cite{}
the two EuO strips induce ferromagnetism as well as {heavy electron doping's}
(with a Dirac point of $E_D=-1.37$eV)
in graphene just underlying them
(see the middle panel of Fig. \ref{fig:setup}).
%required by the spin-dependent transport.
%which
%contains spin-dependent parabolic dispersions and
%heavy electron doping (see the middle panel of Fig. \ref{fig:setup}).
%the ferromagnetism is accompany with
%%%
On top of the EuO strips, two top gates ($V_t^{(i)},i=1,2$) are applied.
They are used to tune the local carrier concentrations and
{to lift up} the Dirac points of the two ferromagnetic graphene's
to align with that of the prinstine graphene
(see the middle panel of Fig. \ref{fig:setup}).
%In this case, spin-dependent barriers can be used. %through an electric field effect of the two gates.
Therefore, %for the chosen system geometry,
%a spin-up and spin-down electron propagating along
%the sample will be subjected to
%two different rectangular barriers. %or wells.
%In other words,
the array of two gated EuO strips %and a proper back gate
creates the desired {ferromagnetic double barriers} or spin-dependent DB-RTD.
Below the substrate, a back gate ($V_g$) is applied to tune the Fermi energy ($E_F$)
through the whole sample.

When a bias voltage ($V_b$) is applied between the source and drain, a net current will be produced
by the electrons
or holes in the source within an energy range
of {$(0,E_F)$ or $(E_F-eV_b,0)$} at zero temperature, see the bottom panel in Fig. \ref{fig:setup}.
Due to the spin difference in the ferromagnetic regions, the current will be spin dependent.
Unlike other works %concerning NDR in graphene
which {considered} a step-like voltage drop through the device,
in this work we also consider a linear voltage drop, $V(x)=f x$ with $f=V_b/l$,
between the source and drain along the $x$-direction.

%%%%%%%%%
%%%%%%%%% begin the formula

The magnetic proximity effect in graphene is usually described by
a simple Zeeman splitting model,\cite{haugen2008spin,munarriz2013strong,diaz2014spin,lu2013tunable,
dell2009wave,niu2008spin,faizabadi2012spin,zou2009negative}
based on which {a} spin wavevector filtering {effect},\cite{lu2013tunable}
{a} spin conductance filtering {effect},\cite{dell2009wave}
and {a} tunneling magnetoresistance {effect}
\cite{niu2008spin,faizabadi2012spin,zou2009negative}
have been proposed.
(These effects have also been found in other two-dimensional systems
subjected to magnetic modulations;
%such as two-dimensional electron gas, surface of a topological insulator, and silicene;
just to cite a few.\cite{wang2002spin,papp2005spin,wu2011spin,shakouri2015tunable,shakouri2014spin})
The Zeeman splitting model
is roughly estimated
from an analogy with a EuO/Al interface
and considers only
opposite energy shifts for opposite spins
added to the linear dispersion of a pristine graphene.\cite{haugen2008spin}
%(i.e., no charge doping is induced).
However, very recent ab-initio calculations\cite{yang2013proximity,hallal2017tailoring} and experiments\cite{swartz2012integration,wang2015proximity,wei2016strong,zhang2017realization}
demonstrate a totally different picture:
the {gapless and linear} Dirac cone for each spin shifts down to a negative Dirac point ($D_s$),
%{the linear dispersion for each spin}
opens a large energy gap ($\Delta_s$), and changes its Fermi velocity ($v_s$)
($s=\pm1$ for spin up and {spin down}),
{all depending on the spin index}; %\cite{song2015spin,song2017electric}
%the change is different for spin.
see middle and bottom panels in Fig. \ref{fig:setup}.
For graphene on six bi-layer EuO, the parameters read
$D_{+(-)}=$42 (-24) meV,
$\Delta_{+(-)}=$67 (49) meV,
and $v_{+(-)}=$ 1.337 (1.628) $v_F$, respectively.\cite{yang2013proximity,song2015spin}
%spin-resolved Dirac cone doping's, %spin-resolved
%Dirac gaps, and %spin-resolved
%Fermi velocities
%large spin-dependent Dirac gaps accompany with parabolic dispersions
%On the other hand, %while
%in their calculations all these works \cite{haugen2008spin,munarriz2013strong,diaz2014spin,lu2013tunable,dell2009wave,niu2008spin,faizabadi2012spin,zou2009negative}
%adopted
%a simple Zeeman splitting model to describe the magnetic proximity effect.
%This model is roughly estimated
%from the analogy with a EuO/Al interface \cite{haugen2008spin}
%and considers only
%opposite energy shifts for opposite spins
%added to the linear dispersion of a pristine graphene.
%%%%
%Since 2012, the magnetic proximity effect have been experimentally realized,
%for example, by depositing EuO \cite{swartz2012integration} or EuS \cite{wei2016strong} on graphene,
%or by transferring graphene onto yttrium iron garnet
%\cite{wang2015proximity,mendes2015spin,leutenantsmeyer2016proximity,evelt2017chiral}
%or La$_{0.7}$Sr$_{0.3}$MnO$_3$ \cite{sakai2016proximity}.
%Interestingly, %a normal ferromagnetism with
%heavy electron doping's and finite Dirac gaps are found
%\cite{swartz2012integration,wang2015proximity,wei2016strong}.
%Ab-initio calculations \cite{yang2013proximity,hallal2017tailoring}
%further demonstrate large spin Dirac gaps accompany with parabolic dispersions
%(see middle panel in Fig. \ref{fig:setup}).
Others\cite{zollner2016theory,su2017effect,hallal2017tailoring} and we\cite{song2015spin,song2017electric}
have developed effective Hamiltonians
in a sublattice-spin direct produce space or a sublattice space,
based on which
nontrivial effects, such as
quantum anomalous
Hall effect,\cite{su2017effect}
simultaneous spin filter and spin valve effect,\cite{song2015spin}
electric-field-induced extremely large {change in resistance},\cite{song2017electric}
and pure crossed Andreev reflection\cite{ang2016nonlocal} have been explored.

%As we have indicated,
%the proximity-induced ferromagnetism in graphene can be viewed as
In a sublattice space{,} the effective Hamiltonian
%{of a ferromagnetic graphene}
reads\cite{beenakker2008colloquium,song2015spin,song2017electric}
%in a sublattice space
\begin{subequations}
\begin{equation}
\mathcal{H}^f_{\bm{k},s,\xi}= \bm\sigma \cdot \hbar v_s \bm{k}+ \xi \sigma_z \Delta_s + \mathcal{I} (D_s +e V_t^{(i)}+e V(x)).\label{Hamiltonian:fm}
\end{equation}
%the ferromagnetism in graphene can be viewed as
Here
%Accordingly,
$\bm{\sigma}=(\sigma_{x},\sigma _{y})$ is a pseudospin Pauli matrices,
$\bm{k}=(k_x, k_y)$ is a momentum operator,
$\xi =\pm 1$ is an index for valley $K$ and $K^\prime$,
and $\mathcal{I}$ is an {identity} matrix.
%and $V(x)=xV/l$ is the voltage drop.
For the pristine graphene in the well and {the} buffer regions,
%or between the ferromagnetic ones,
the Hamiltonian is well known
\begin{equation}
\mathcal{H}^p_{\bm{k}}= \bm\sigma \cdot \hbar v_F \bm{k} + \mathcal{I} e V(x).\label{Hamiltonian:pr}
\end{equation}
For the contacted graphene in the source and drain{,} the Hamiltonian reads
\begin{equation}
\mathcal{H}^c_{\bm{k}}= \bm\sigma \cdot \hbar v_F \bm{k} + \mathcal{I} U.\label{Hamiltonian:con}
\end{equation}
\end{subequations}

{For convenience, below we will express all physical quantities in their dimensionless
form,} in terms of a characteristic length $l_0=$56.55nm and
a corresponding energy unit $E_0=\hbar v_F/l_0=$10meV.
%Under this condition the charge density of
%states is coinciding with a true system and
%a coherent transport regime is ensured even at room temperature.
The transversal momentum $k_y=(E-U)\sin\alpha$
($E$ and $\alpha$ are the energy and incident angle of a fermion, respectively)
% together with the wave function $e^{ik_yy}$
is conserved, and it suffices to {solve}
the longitudinal wave function $\bm{\Psi}_j(x)$,
where $ j=c,f,p$ for the contacted, ferromagnetic, and pristine graphene,
respectively.
However, straightforward decoupling of $\mathcal{H}_j\bm\Psi_j(x) = E \bm\Psi_j(x)$ in Eqs. (\ref{Hamiltonian:fm}) and (\ref{Hamiltonian:pr}),
which contain a linear term of $x$,
results in an unsolvable two-order differential equation.\cite{sonin2009effect}
To solve this problem, we perform a rotation
of the Dirac equation by $\pi/2$ around the $y$-axis in the
pseudospin space (see the upper pannel of Fig. \ref{fig:setup}).
The envelope functions become resolvable and the result
for the ferromagnetic graphene reads
\begin{subequations}\label{wavefunction}
\begin{equation}\label{wavefunction1}
\bm{\Psi}^s_f=
f^+\left(\begin{array}{c}
F_s(x)\\
G_s^*(x)\end{array}\right)e^{ik_{y}y}
+f^-\left(\begin{array}{c}
G_s(x)\\
F^{*}_s(x)\end{array}\right)e^{ik_{y}y},
\end{equation}
% recurrence relation
where $F_s(x)=D[-1/2+iq_s^2/2a_s,(1+i)(E-D_s-eV_t^{(i)}+fx)/\sqrt{a_s}]$ and
$G_s(x)=D[-1/2-iq_s^2/2a_s,(-1+i)(E-D_s-eV_t^{(i)}+fx)/\sqrt{a_s}]$.
{Here $D[,]$ is} the Weber parabolic cylinder function,
$q_s^2=\Delta_s^2+k_y^2$, and $a_s=f v_s$.
Note, $F_s(x)$ and $G^{*}_s(x)$ [$F^{*}_s(x)$ and $G_s(x)$] are spin
resolved and have properties of a right (left)-going wave function.\cite{sonin2009effect}
%%%
For the pristine graphene %(p-n junction),
the result becomes %\cite{song2013negative}
\begin{equation}\label{wavefunction2}
\bm{\Psi}_p=p^+\left(\begin{array}{c}
F(x)\\
G^{*}(x)\end{array}\right)e^{ik_{y}y}
+p^-\left(\begin{array}{c}
G(x)\\
F^{*}(x)\end{array}\right)e^{ik_{y}y},
\end{equation}
where
$F=D[-1/2+ik_y^2/2f,(1+i)(E+fx)/\sqrt{f}]$ and
$G=D[-1/2-ik_y^2/2f,(-1+i)(E+fx)/\sqrt{f}]$.
In the rotated pseudospin space, the envelope functions for the contacted graphene
become
\begin{equation}\label{wavefunction3}
\bm\Phi_c^\pm=
c^\pm\left(\begin{array}{c}
%\frac{E_j\pm k_j+iq_j}{2E_j}
1\pm e^{\pm i \alpha} \\
-1 \pm e^{\pm i \alpha}
\end{array}\right)
%[E_j\pm k_j+iq_j,-E_j\pm k_j+iq_j]^T
e^{\pm ik_cx+ik_yy},
\end{equation}
\end{subequations}
where $k_c^{s,d}=\textmd{sign} [E-U+eV_b^{s,d}]\sqrt{(E-U+eV_b^{s,d})^2-k_y^2}$
with $V_b^{s,d}=0,V_b$.
In the {device} source, $c^+=1$ and $c^-=r_s$,
while in the {device} drain $c^+=t_s$ and $c^-=0$.
$r_s$ and $t_s$ are spin-dependent reflection and transmission coefficients, respectively.

\begin{figure}[!t]
  \centering
  \includegraphics[width=0.95\linewidth]{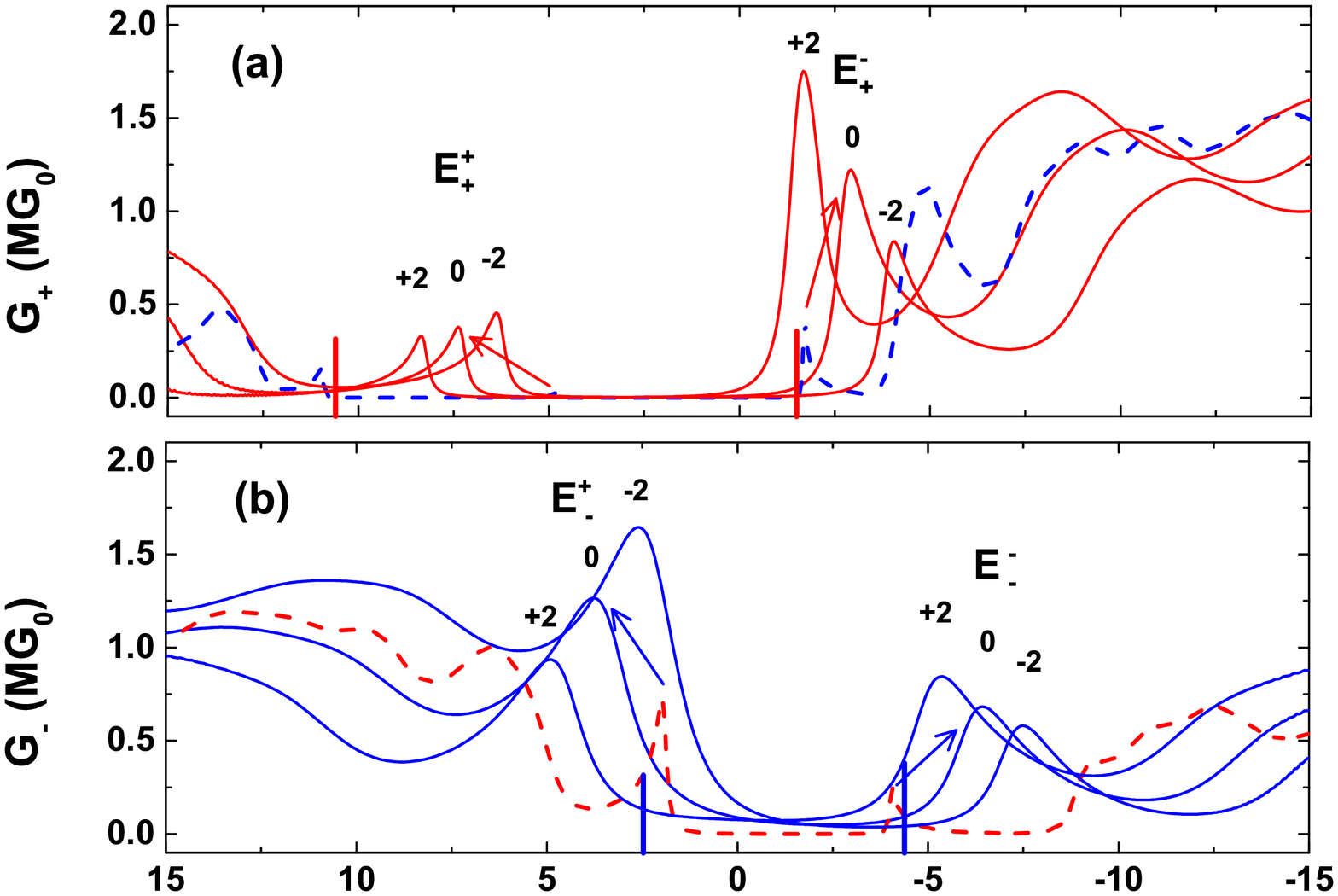}\\
  \includegraphics[width=0.95\linewidth]{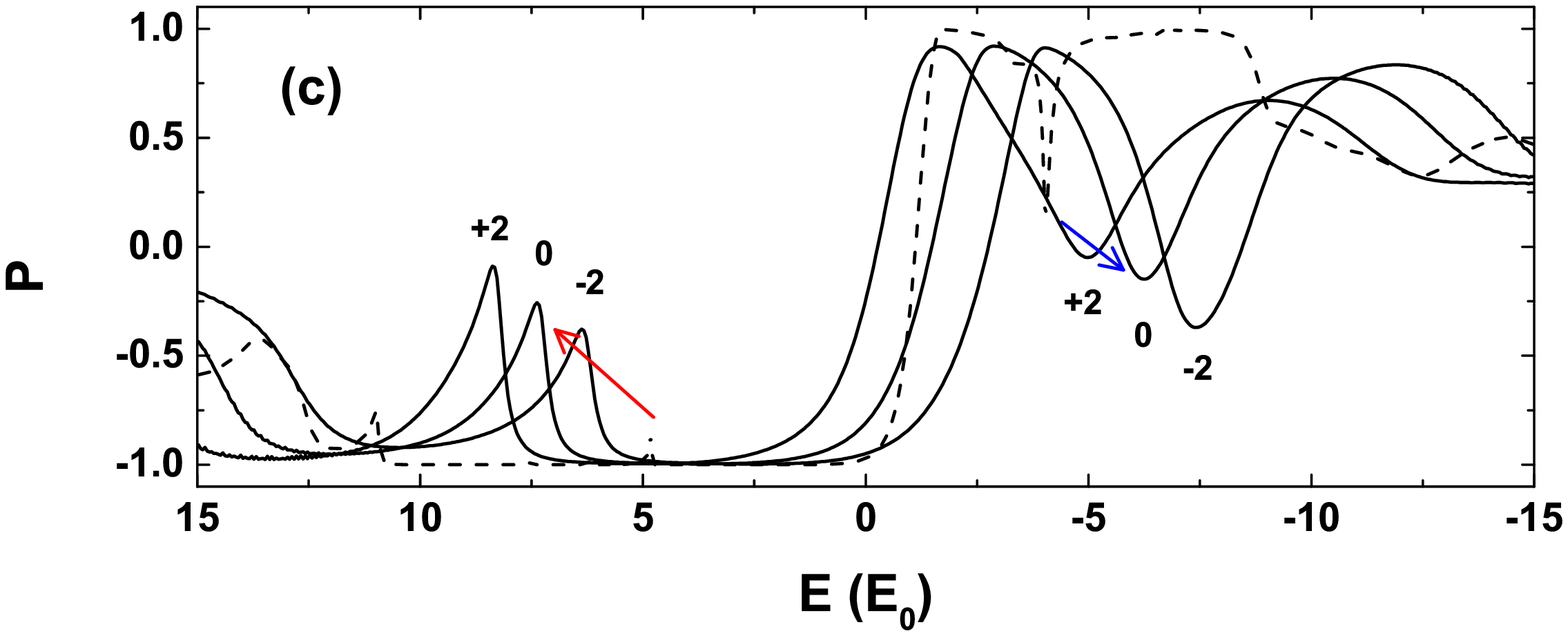}
  \caption{(color online) (a,b) Linear conductances for spin up (a) and spin down (b)
 as a function of the Fermi energy.
Dashed: $l=2$ and $eV_t^{(i)}=0$,
solid: $l=1$ and $eV_t^{(i)}=-2,0,2$.
  (c) The corresponding spin polarizations as a function of the Fermi energy.
  {An ideal electrode doping $U=0$ is considered.}
  % there are 6 device parameters: l, Vt, U; EF, Vb, spin.
  }\label{fig:unbias}
\end{figure}

With the standard transfer-matrix method,\cite{born2013principles}
$t_s$ can be obtained by matching the
spinor envelope functions in {Eqs. (\ref{wavefunction1}), (\ref{wavefunction2}), and (\ref{wavefunction3})}
at the potential boundaries.
%($x=0,0.1a,1.1a,1.1a+b,2.1a+b,l$).
The {spin-dependent} transmission probability reads
$
T_s(E,\alpha,V_b)=|t_s|^{2}
(k_c^d/E-U+eV_b)(k_c^s/E-U)^{-1}
$
for $(k_c^s)^2>0$ and $(k_c^d)^2>0$,
and $T_s=0$ otherwise.
The spin-dependent net current at zero temperature can be calculated by the
Landauer-B\"{u}ttiker formalism\cite{buttiker1985generalized}
\begin{equation}\label{eq:current}
I_s(E_F,V_b)=I_0\int_{E_{F}-eV_b}^{E_{F}}\int_{-\pi/2}^{\pi/2}T_s (E,\alpha,V_b)
|{E-U}|
\cos\alpha
d\alpha dE,
\end{equation}
where $I_0=2ev_{F}W/(2\pi l_0)^2$ is a current unit with
the factor 2 coming from the valley degeneracy.
Considering the values of $W$ and $l_0$, $I_0=60\mu A$,
which is one or two orders of magnitudes larger than
those in {devices based on} graphene nanoribbons.\cite{wu2012negative,jiang2014spin,yan2014effects,yang2015half,xu2014negative,
hong2015axisymmetric,son2006half,liu2016efficient,munarriz2013strong}
%The total current and spin polarization are defined as
%$I=I_++I_-$ and $P_I=(I_+-I_-)/I\times 100\%$.
$(E_F-eV_b,E_F)$ is an energy {integral interval}, in which
a spin-dependent differential conductance at a given energy can be defined as
$g_s(E,V_b)\equiv\int_{-\pi/2}^{\pi/2}T_s(E,\alpha,V_b) \cos\alpha d\alpha$.
It is also seen that, $g_s(E,V_b)$ contributes to $I_s(E_F,V_b)$ by a weight of
$M(E,W)=2|{E-U}|W/hv_F$, which is
%The integrated weight of $|E|$ together with $W$ in $I_0$
%in the energy integration,
% it contributes to the current under bias by an integrated weight of $|E|$,
the transversal modes number (TMN) at $E$.\cite{tworzydlo2006sub}

\section{Results and Discussion}\label{sec:III}

\subsection{Spin-dependent linear conductance at zero bias}\label{sec:III.A}

Figure \ref{fig:unbias} (a,b) {show} the spin-dependent linear conductance through
the unbiased device as a function of the Fermi energy:
$G_\pm(E_F,V_b=0)=M G_0\int_{-\pi/2}^{\pi/2}T_s(E_F,\alpha,V_b=0) \cos\alpha d\alpha$,
where $G_0=e^2/h$ is the quantum conductance. %(two accounts for the valley degeneracy)
%and $M=2|E_F|W/hv_F$ is the number of the transversal modes.
The calculations can be made by following Ref. \cite{song2017electric}.
{Here an ideal contacting ($U=0$) is considered.}
It is seen that, already for a relatively short device ($l=2$ with $a\rightarrow2a$ and $b\rightarrow2b$, the dashed curve),
its linear conductance for each spin shows a good accordance to the corresponding band gap,
%shown in the middle panel of Fig. \ref{fig:setup}.
whose conduction band minimum (CBM) and valence band maximum (VBM) are
indicated by the vertical bars {in Fig. \ref{fig:unbias} (a,b)}.
Inside each {energy} gap, %the conductance is rather small.
two resonant peaks $E_+^\pm$ and $E_-^\pm$ arise, {where
the subscript $\pm$ stand for spin up and spin down and the superscript $\pm$
stand for the CBM or VBM part}.
This is a clear significant of resonant tunneling in the device. %between the two barriers.
For a shorter device ($l=$1, the solid {curves}) the resonant peaks shift outside
and become stronger (indicated by the arrows).
The spin-dependent linear conductance at different top gates are also shown.
It is seen that, the resonant levels become larger (smaller)
as a positive (negative) top gate is applied.
This will play an important role in the gate control of the spin-dependent NDR.
{Compared} with spin up, the resonant tunneling is much weaker for spin down.
This is a result of a smaller effective spin Dirac gap,
i.e., $\Delta_\textmd{eff}^-=\Delta_-/v_-\approx 3
<\Delta_\textmd{eff}^+=\Delta_+/v_+\approx 5$.\cite{song2017electric}

The total conductance is spin dependent, which manifests itself clearly in the
polarization, which is defined as $P_G=(G_+-G_-)/(G_++G_-) \times 100\%$ and plotted {in Fig. \ref{fig:unbias}(c)}.
{In Fig. \ref{fig:unbias}(c)},
near $-100\%$ ($100\%$) polarization can be found
in the {energy} ranges between CBMs (VBMs) of different spins.
%i.e., the electron (hole) splitting window.
They are much higher than spin polarizations
obtained in Refs. \cite{dell2009wave,zou2009negative}
and can be applied as a highly efficient spin conductance filtering.
On the other hand, %unlike them
the spin filtering effect is found to be weaken by the resonant tunneling
and can be exactly controlled by the top gates.

\begin{figure}[!t]
\centering
\includegraphics[width=0.95\linewidth]{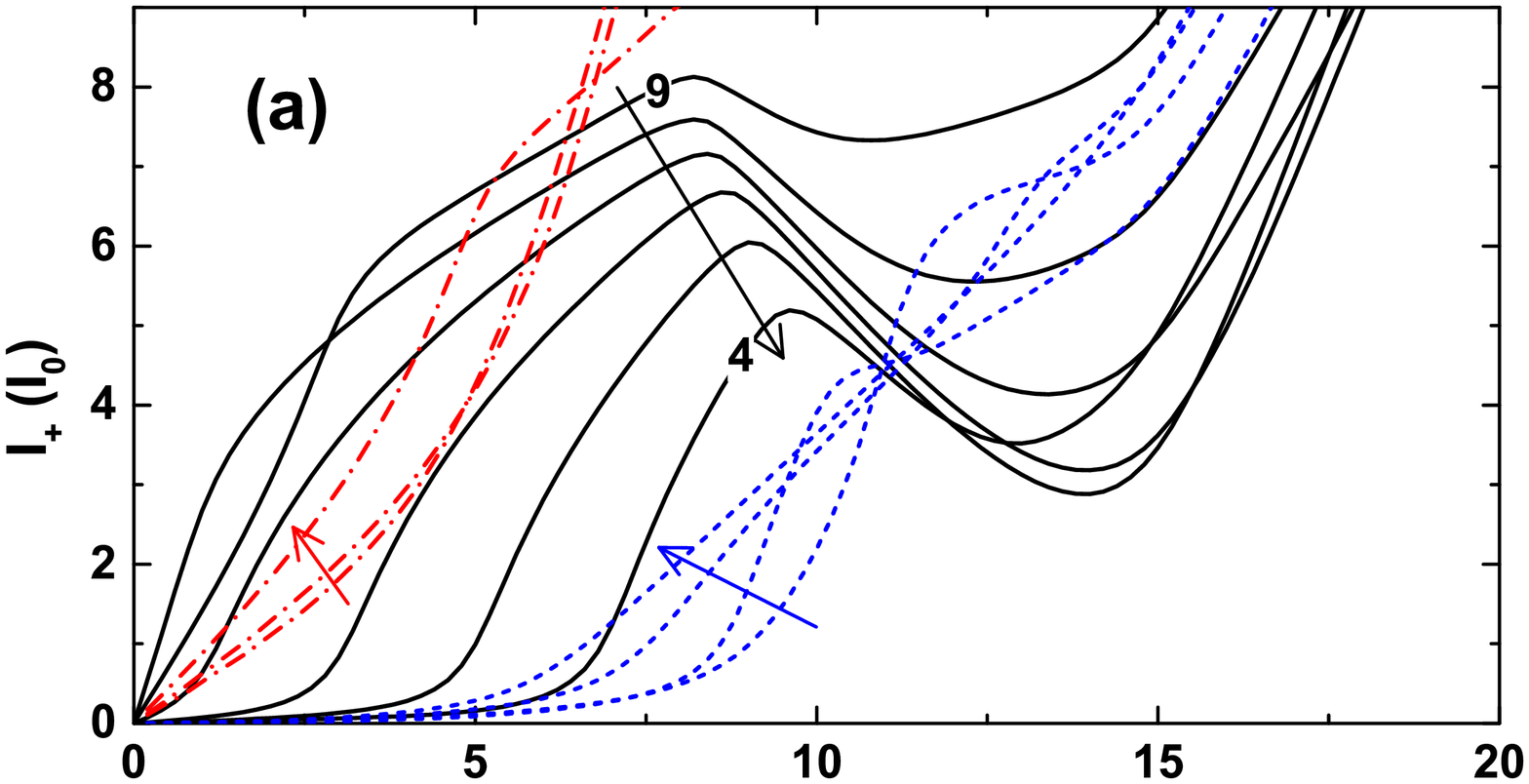}\\
\includegraphics[width=0.95\linewidth]{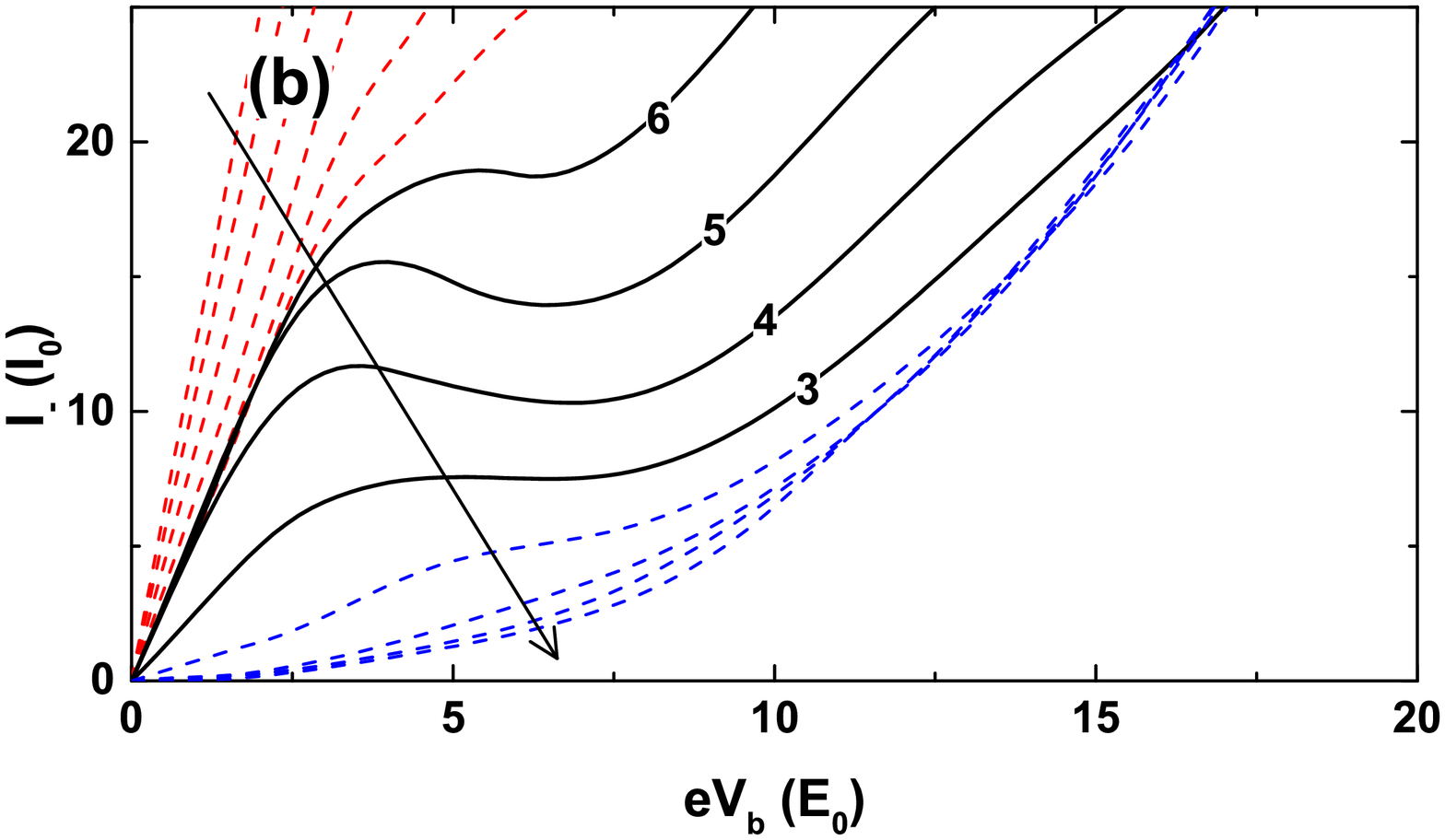}
\caption{(color online) %(a, b) Spin-dependent differential conductance as a function of the energy for several typical biases.
I-V characteristics for (a) spin up and (b) spin down at various Fermi energies with a step of 1.
{In (a), $E_F$ changes
from 12 to 10, from 9 to 4, and from 3 to 0 for curves along the red, black, and blue arrows, respectively.
In (b), $E_F$ decreases from 12 to 0 for curves along the arrow.
In both (a) and (b), the curves with no NDR features are plotted in dashed.
The other device parameters are
$l=1$, $eV_t^{(i)}=0$, and $U=0$.}
%(e, f) Total current and its spin polarization as a function of the bias for various Fermi energies indicated in the figure.
}\label{fig:biased}
\end{figure}

\subsection{Spin-dependent NDR at finite bias}\label{sec:III.B}

%%%
The spin-dependent I-V {characteristics} of the device %currents as a function of bias
{with} various Fermi energies are plotted in Fig. \ref{fig:biased}(a,b). %and \ref{fig:biased}(d).
%Here are considered.
It is seen in Fig. \ref{fig:biased}(a) that, for Fermi energies around the resonant level $E_+^+=7.37$
(black solid), the {spin-up} current displays obvious NDR.
However, for Fermi energies much higher ($>9$, red dashed)
or smaller ($<3.5$, blue dashed) than the resonant
{level}, %the resonant peak,
the NDR features disappear.
The case is the same for spin down as shown in Fig. \ref{fig:biased}(b), where NDR features
{are found only
for} Fermi energies around the resonant level $E_-^+ = 3.77$.

\begin{figure}[!t]
\centering
\includegraphics[width=0.95\linewidth]{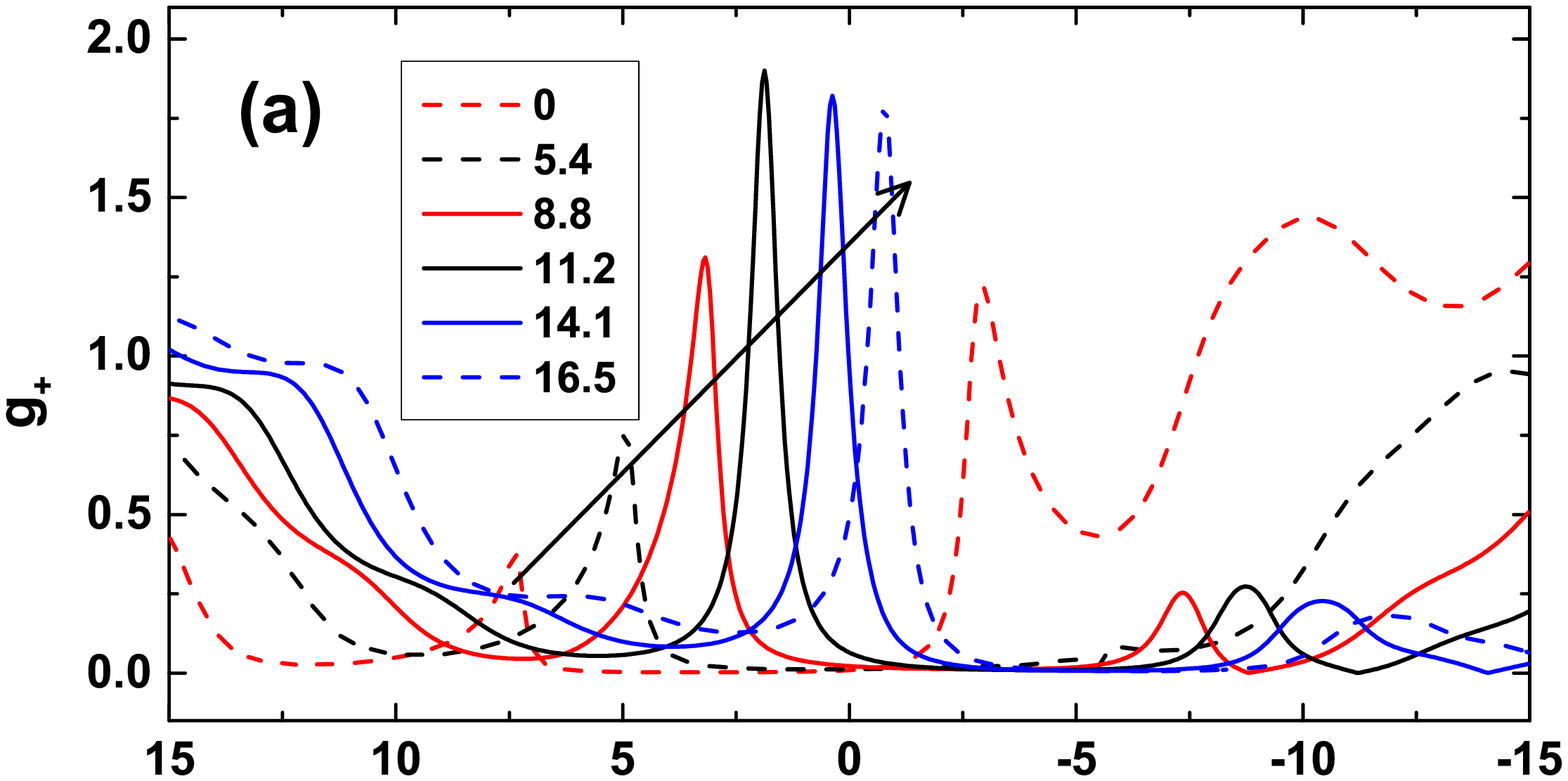}\\
\includegraphics[width=0.95\linewidth]{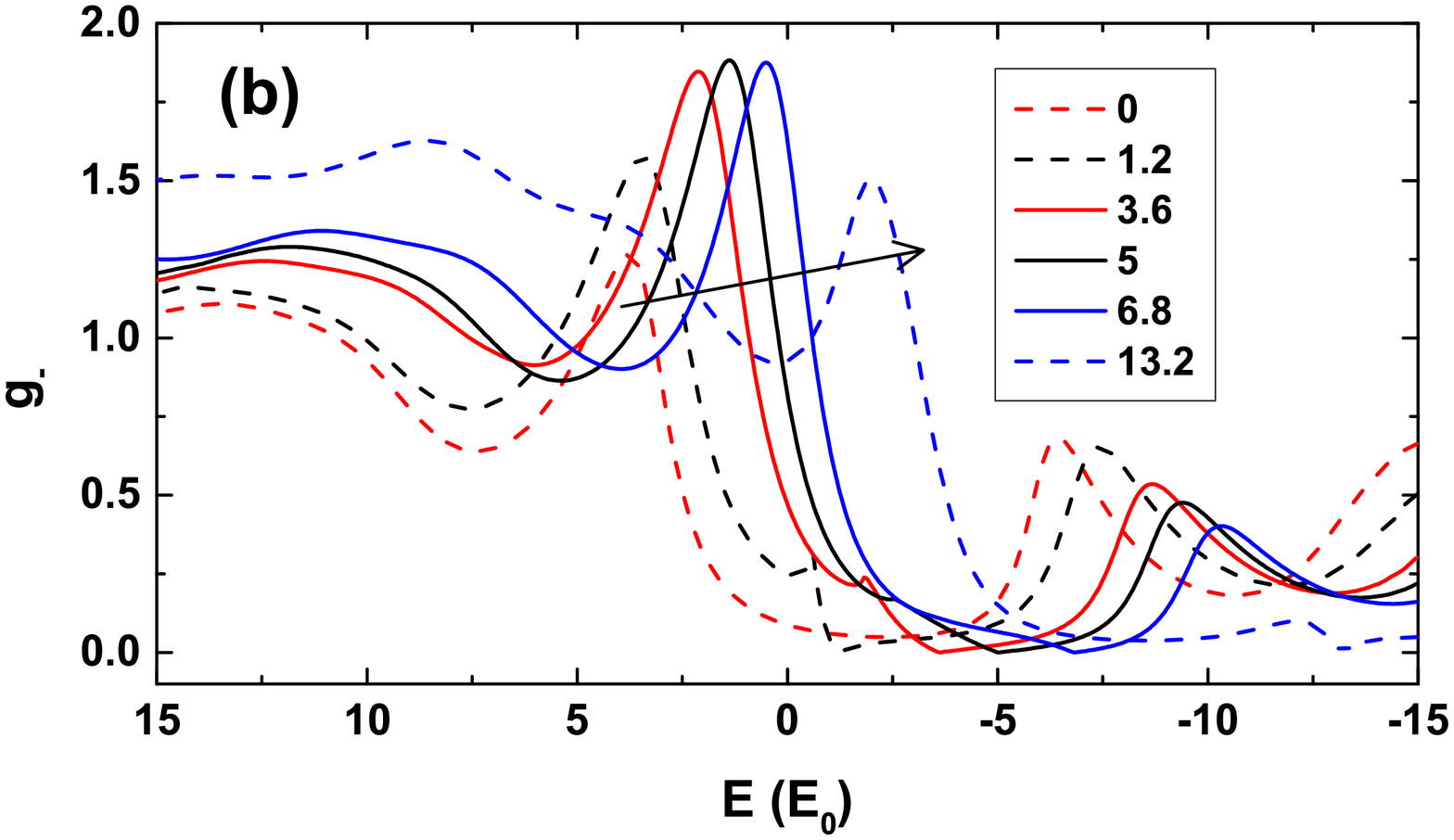}
\caption{(color online) Differential conductance as a function of the energy for (a) spin up and (b) spin down
at several typical biases. %indicated in the figure.
{Curves with biases in the NDR region are plotted in solid: red for current peak,
blue for current valley, and black for between.
Curves with biases out of the NDR region are plotted in dashed: red for zero bias,
black for bias before the current peak, and blue for bias after the current valley.
The other device parameters are $l=1$, $eV_t^{(i)}=0$, and $U=0$.}
%(c, d) I-V characteristics at various Fermi energies with a step of 1 for spin up and spin down.
%(e, f) Total current and its spin polarization as a function of the bias for various Fermi energies indicated in the figure.
}\label{fig:dc}
\end{figure}

%\begin{figure}[!t]
%\centering
%%\includegraphics[width=\linewidth]{control}
%\includegraphics[width=0.5\linewidth]{both}
%\caption{
%%(a, b) Spin-dependent I-V characteristics for $E_F=5.5$ and various top gates.
%%In (a) the top gates increase from -6 to 4 (6 to 14) with a step of 2
%%along the red (blue) arrow;
%%In (b) the top gates increase from -6 to 0, 2 to 8, and 10 to 14 with a step of 2, respectively.
%(c) The bias voltages for the current peak and valley (solid) and the PVR (dashed) as a function of the top gates,
%red for spin up and blue for spin down.
%(d) Dual-spin NDR at top gates of 0 and 2.
%}\label{fig:control}
%\end{figure}

\begin{figure*}[!t]
\centering
\includegraphics[width=0.49\linewidth]{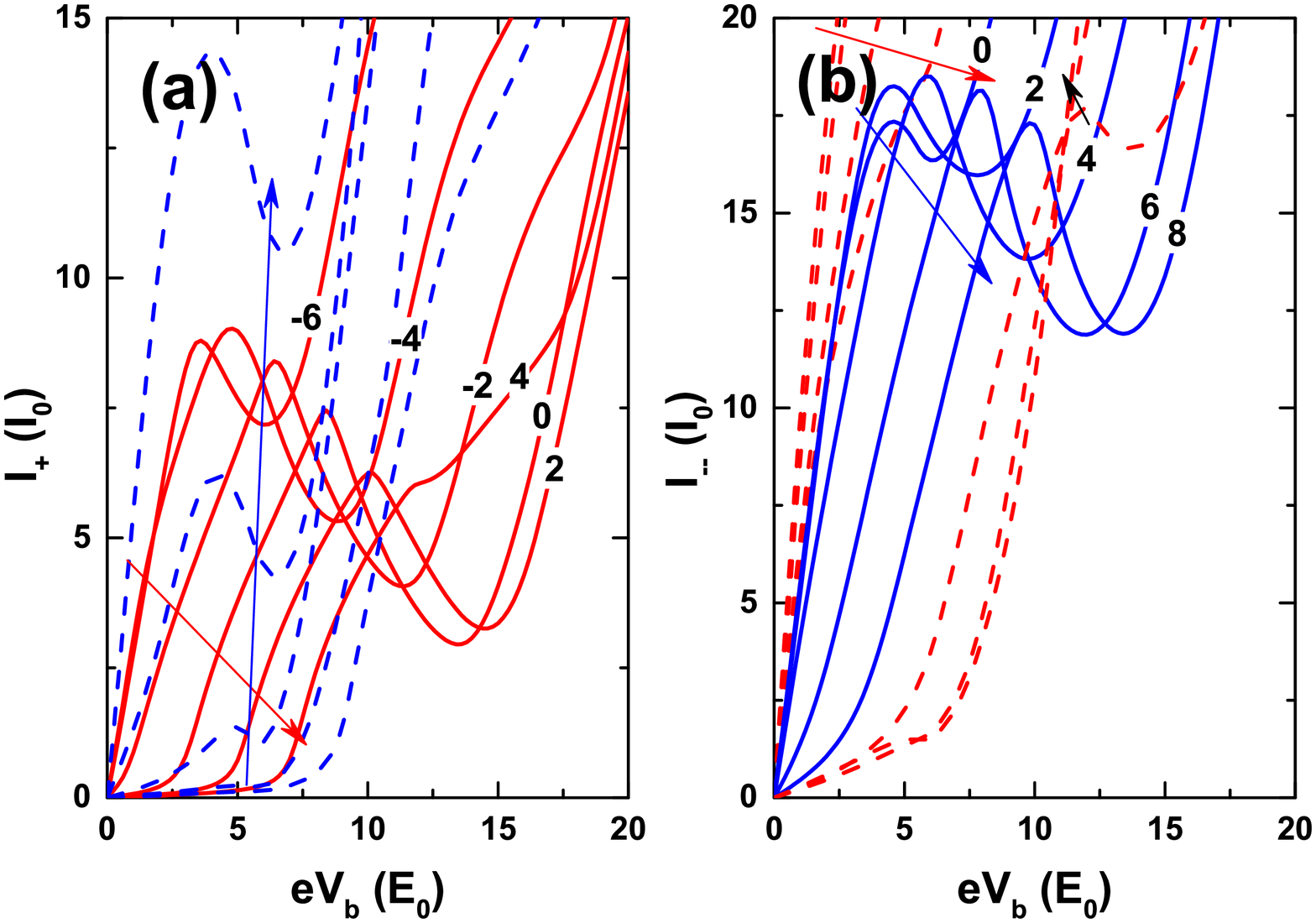}
\includegraphics[width=0.49\linewidth]{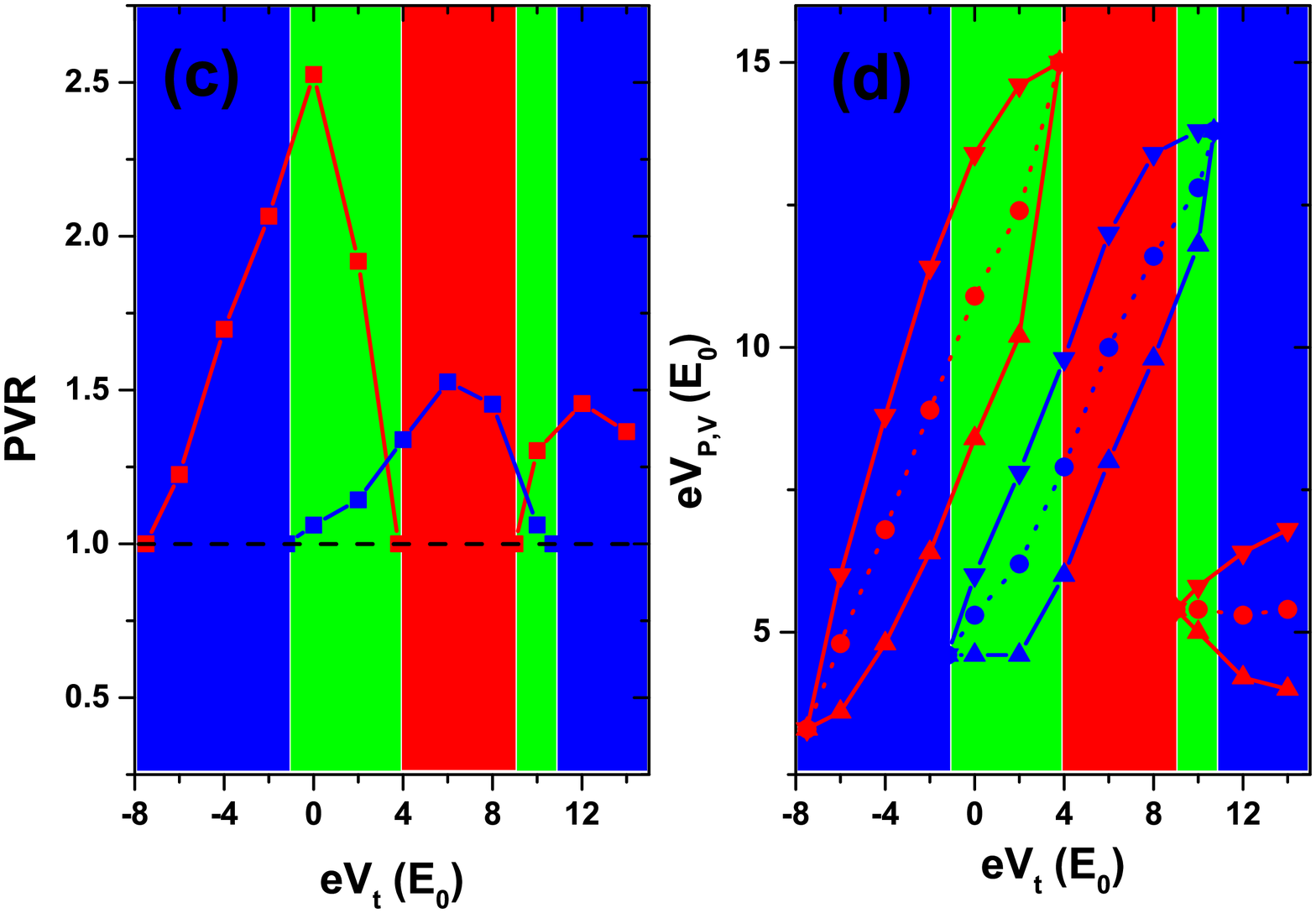}
\caption{(color online) (a, b) Spin-dependent I-V characteristics for $E_F=5.5$ at various top gates.
In (a) the top gates increase from -6 to 4 (6 to 14) with a step of 2
along the red (blue) arrow;
In (b) the top gates increase from -6 to 0, 2 to 8, and 10 to 14 with a step of 2 along the red, blue, and black arrow, respectively.
{(c) The PVR as a function of the top gates
for spin up (red) and spin down (blue).
(d) The NDR {region} (upward triangles for $eV_P$ and
downward triangles for $eV_V$)
as a function of the top gates
for spin up (red) and spin down (blue).
%The dual-spin NDR at top gates of 0 (solid) and 2 (dashed), red for spin up and blue for spin down.
The other device parameters are $l=1$ and $U=0$.}
}\label{fig:control}
\end{figure*}

To understand these interesting behaviors, we plot the {spin-up} differential conductance
as a function of energy in Fig. \ref{fig:dc}(a). %and \ref{fig:biased}(b).
Several typical biases in the strongest NDR case ($E_F=5.5$) are considered.
It is seen that, for Fermi energies around the resonant level,
the transport under the considered biases are dominated by the resonant state.
%%%%%
As the bias increases from zero (the $eV_b=$ 0 and 5.4 {curves}), %the integrated window enlarges while
the resonant peak shifts into
the {integral interval} (the origin of which, $E_F$, is indicated by the vertical bar)
with an increasing peak value.
This {results in} an increase behavior of the spin-up current with an increasing bias in the device.
The {spin-up} current reaches its maximum at a bias of $eV_P=8.8$,
where the resonant peak has totally inserted into the {integral interval}.
As indicated above,
%It is seen from Eq. (\ref{eq:current}) that,
the spin-up resonant state contributes to the spin-up current by a
weight of TMN$\propto|E|$ {(where $U=0$)}, see Eq. (\ref{eq:current}).
%, which is proportional to an absolute value of its energy
%which is similar to a density of state.
As the bias increases further (the $eV_b=$11.2 {curve}), the TMN decreases,
which results in a decrease behavior of the {spin-up} current with an increasing bias, i.e., the desired NDR feature.
The TMN decreases {with the bias} because, as the bias increases
the \emph{positive} resonant level {is pulled down}
and gets closer to the Dirac point in the source, i.e., $|E|=E$ becomes smaller
as $E$ becomes smaller.
The spin-up current achieves its minimum at a larger bias ({the}
$eV_V=14.1$ {curve}),
where the {pull-down} resonant level almost aligns to the Dirac point in the source %($E=0$)
and the TMN approaches {near} zero. %$V_V\sim 2E^+$.
However, as the bias increases further (the $eV_b=$16.5 {curve}),
the {pulling down} of a \emph{negative} resonant level leads to an increase of the TMN,
i.e., $|E|=-E$ becomes bigger as $E$ becomes smaller.
As a result, the {spin-up} current increases again. %with an increasing bias.

On the other hand,
for Fermi energies far from the {spin-up} resonant level, the biased transport is also contributed
by {a} quasi-ballistic state outside the spin Dirac gap, see Fig. \ref{fig:dc}(a).
As a result, the bias-induced decrease of the TMN ({integral weight}) %of the resonant level
is counteracted by the bias-induced increase of the {integral interval},
and the NDR feature disappears.
%%%%
For spin down, the NDR is much less pronounced.
%NDR are found for different Fermi energies
%around $E_-^+ = 3.77$ (see Fig. \ref{fig:biased}(b)).
This stems from a more expanding resonant peak, which is shown in Fig. \ref{fig:dc}(b),
where several typical biases for the strongest NDR case ($E_F=4.0$) are considered.

\subsection{Top-gate control of the spin-dependent NDR}

%We now consider
The spin-dependent I-V characteristics at various top gates
are shown in Fig. \ref{fig:control}(a,b).
Here $E_F=5.5$ is focused, for which both spins display obvious NDRs.
It is seen that, as the top gates increase in the considered range,
spin up displays NDR (the -6 to 2 {curves}), no NDR (the 4 to 8 {curves}), and NDR (the 10 to 14 {curves}), respectively.
In contrast, spin down shows no NDR (the -6 to -2 {curves}), NDR (the 0 to 10 {curves}), and no NDR
(the 12 to 14 {curves}), respectively. %as the gates increase.
This is according to that the resonant levels $E_+^+$, $E_-^+$, and $E_+^-$ are
sequentially {lifted up} around the Fermi energy and {dominate} the biased transport.
%In this case, the resonant state $E_-^+$ is lift around the Fermi energy and utilized.
%%%
For each spin, the peak-to-valley currents ratio (PVR)
%and NDR windows ($V_P^\pm,V_V^\pm$) %and ($V_P^-,V_V^-$),
are summarized in Fig. \ref{fig:control} (c) as a function of the top gates.
%The spin-dependent
%as well as PVRs are summarized in Fig. \ref{fig:control}(c).
%From the figure, the spin dependence of the NDR becomes clearer.
%Obvious NDR features are also found for spin up around $E_+^-=-2.93$
%and for spin down around $E_-^-=-6.43$.
It is seen that, according to the occurrence of the spin-dependent NDR
only at certain top gates,
%when the Fermi energy is around the corresponding resonant level,
the PVR shows a non-monotonous dependence on the top gates,
increasing from 1 (no NDR) to certain value and then decreasing to 1 again.
The maximal PVR for each spin is found
at certain top gates that the lifted resonant level is slightly lower than the Fermi energy.
%%%
For spin up, the value reads 2.5, which is comparable with those
obtained in experiments on spin-independent NDRs.\cite{britnell2013resonant,sharma2015room}
%We also find that, a certain asymmetry of the barriers ($\Delta V_t=V_t^{(1)}-V_t^{(2)}<0$)
%can enhance the PVR.
%For $\Delta V_t=-2$ the value changes from 2.18 to 2.52.

Although the NDR happens at different top gates for spin up and spin down,
%where their resonant levels are lift around the fixed Fermi energy,
NDRs for both spins can be achieved for
certain top gates, {see the green windows in Fig. \ref{fig:control} (c)
and the $eV_t^{(i)}=0,2$ curves in Fig. \ref{fig:control}(a,b).
%This is also clear in ,
%in which the cases are plotted. %show NDR for both spins.
In Fig. \ref{fig:control} (a,b),} it is also observed that the NDR {regions}
for different spins do not overlap.
As a result, %for Fermi energy between 3.5 and 6
the proposed device can be applied as
\emph{a {dual-spin} Esaki diode}: at low {biases} spin down displays NDR and
at high {biases} spin up displays NDR.
The {dual-spin} NDR occurs because the resonant levels are lifted up that
the Fermi energy lies between $E_+^+$ and $E_-^+$ ($E_-^+$ and $E_+^-$).
%In another view, this remarkable behavior adds spin degree of freedom to multiple separate NDR regions.
For top gates higher (lower) than the left green {window in Fig. \ref{fig:control}(c)}, i.e., the red (blue) windows,
only spin down (spin up) shows NDR
while the current for the other spin increases monotonously with the bias.
This can be applied as \emph{a {single-spin} Esaki diode}.
Hence, as the top gates
increases from -7 to 15,
the proposed device works consecutively as
a spin-up only, dual-spin, spin-down only, dual-spin, and spin-up only
Esaki diode.
In other words, spin-dependent NDR with spin selectivity can be realized in the proposed device,
by simply {changing} the top gates.
%From the five windows in Figs. \ref{fig:control} (c) we can see that,
%which is similar to that happens under increasing Fermi energies (see Fig. \ref{fig:PVR}).

\begin{figure*}[!t]
  \centering
  \includegraphics[width=0.49\linewidth]{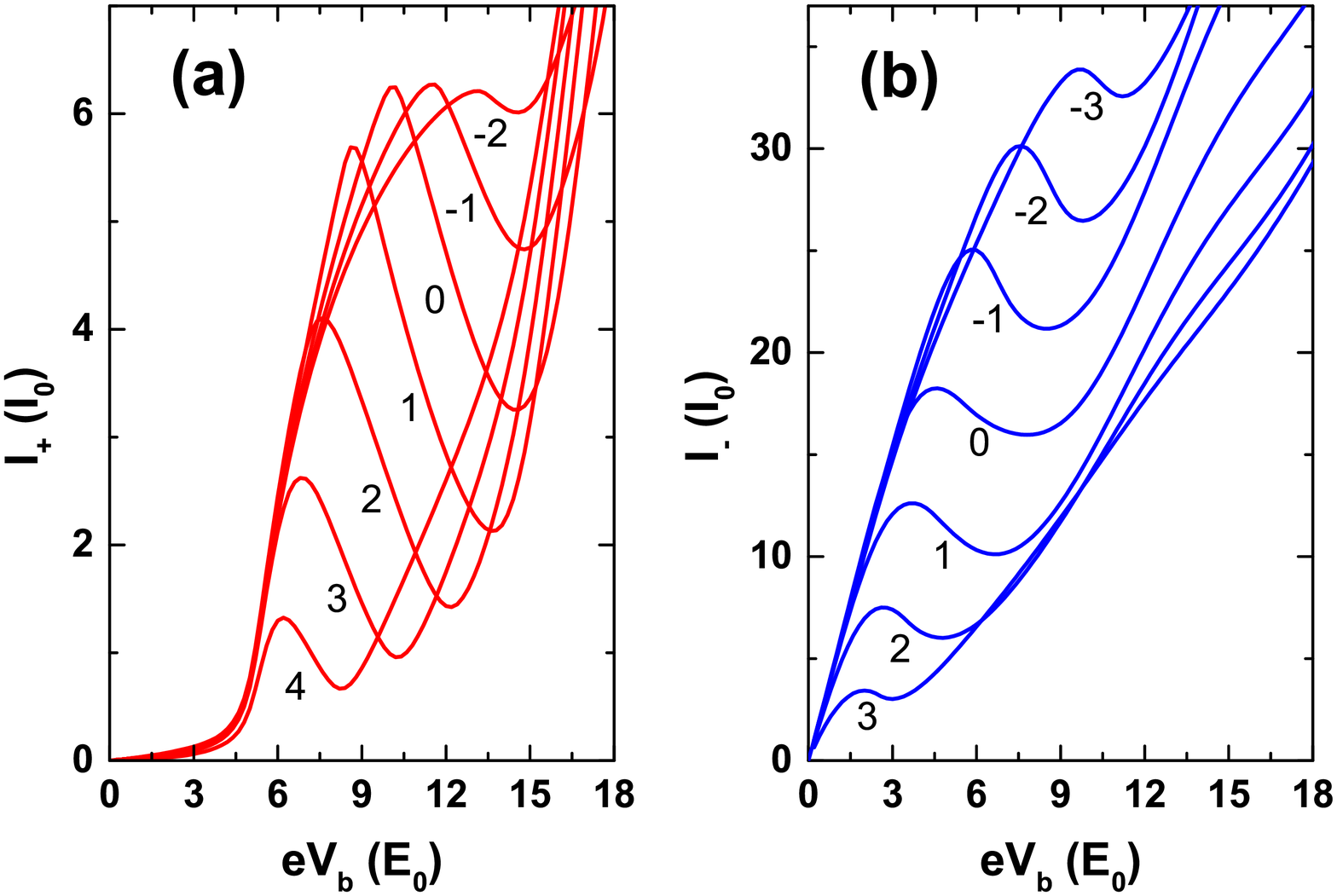}
\includegraphics[width=0.49\linewidth]{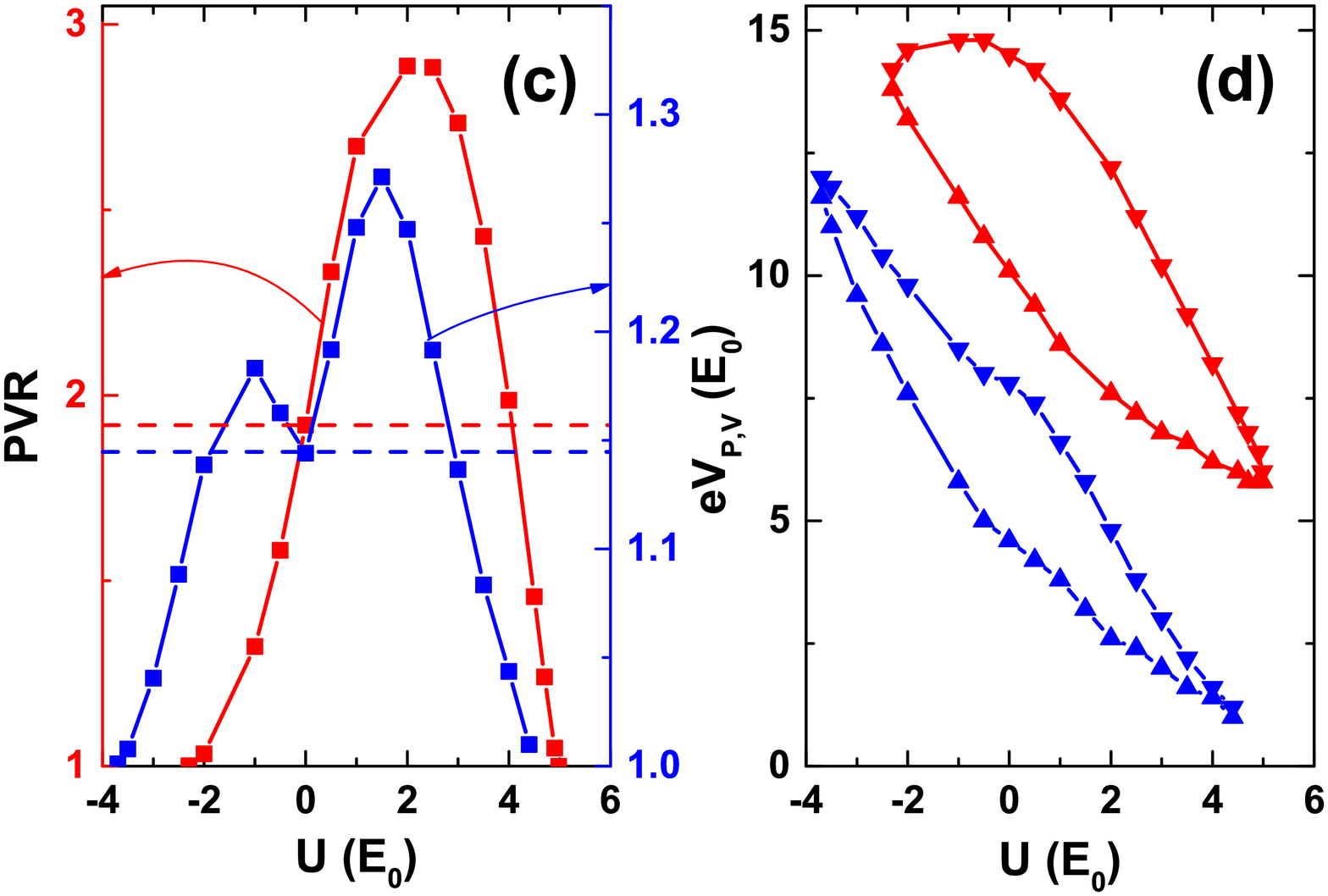}\\
  \caption{{(color online) (a) Spin up and (b) spin down I-V characteristics at different electrode doping's indicated below the curves.
%along the arrow the doping increases with a step of 1.
(c) PVR for spin up and spin down (red and blue) as a function of the electrode doping.
(d) Biases for peak and valley currents (upward and downward triangles) of spin up and spin down (red and blue) as a function of the contact doping.
%  (b) Peak and valley currents (red solid, upward and downward triangles) and PVR (blue solid) for spin up at different contacting;
%  PVR for spin down (blue dashed) at different contacting.
The other parameters are $l=1$, $E_F=5.5$, and $V_t^{(i)}=2$.}}\label{fig:electrode}
\end{figure*}

The NDR {regions} ($eV_P^\pm,eV_V^\pm$) for each spin %and ($V_P^-,V_V^-$),
{are summarized in Fig. \ref{fig:control} (d) as a function of the top gates}.
It is seen that, when the top gates increase from -7.5 to -1.2
(the left blue window in Fig. \ref{fig:control} {(d)}),
the center of the {spin-up NDR region} shifts monotonously %to a higher bias.
from $\sim$3.3 to $\sim$9.7.
%The lowest window centers at $\sim$3 and the highest window centers at $\sim9$.
This implies that, %by tuning the top gates %in this range
the operation {region} of the spin-up Esaki diode
is variable in a large range of $\sim$ 64meV. %can change in.
When $eV_t^{(i)}$ increase from -1.2 to 3.7 (the left green window in Fig. \ref{fig:control} {(d)}),
both the {spin-up and spin-down NDR regions} of the dual-spin Esaki diode are tunable.
%the centers of NDR windows for both spin up and spin down shift monotonously.
%from $\sim$9.7 to $\sim$15 and from $\sim$4.5 to $\sim$7.8.
%to a higher bias.
%In other words, the two.
%can be exactly controlled.
The variable ranges are $\sim53$ meV and $\sim33$ meV, respectively.
For top gates changed between 3.8 and 9 (the red window in Fig. \ref{fig:control} (d)),
the operation {region} of the spin-down Esaki diode is monotonously tunable
in a range of $\sim$45 meV.

The spin selectivity and {region} tunability of the spin-dependent NDR
can be understood as follows.
As the top gates increase, the spin-dependent resonant levels formed
between the {ferromagnetic double barriers} are lift.
As a result, $E_+^+$, $E_-^+$, and $E_+^-$ %, and $E_-^-$
are consecutively closed to the fixed Fermi energy hence
{dominate} the biased transport, solely or simultaneously. %as the top gates increase.
The former leads to a {single-spin} Esaki diode while the latter
results in a {dual-spin} Esaki diode.
On the other hand, for each resonant level dominating the {biased} transport,
its {deviation} from the Dirac point in the source enlarges with the increasing top gates,
see Fig. \ref{fig:unbias}(a,b).
From the NDR mechanism discussed in Sec. \ref{sec:III.B},
%As a result,
the decrease of the TMN ({integral} weight) %of the spin-dependent resonant level
hence the spin-dependent NDR should be happen at a higher bias.
In Fig. \ref{fig:biased} the spin-dependent NDR {region is} also found to
be controlled by the Fermi energy.
However, the change is non monotonous.

%\begin{figure}[!t]
%  \centering
%  \includegraphics[width=\linewidth]{pvr-contacting}\\
%  \caption{{(color online)
%%(a) Spin up I-V characteristics at different electrode contacting.
%}
%}\label{fig:pvr}
%\end{figure}

Spin oscillator, spin amplifier, spin switching, and spin memory
%for a certain spin
can be realized based on a spin-dependent NDR.
The gate-induced spin selectivity in the proposed device adds a spin degree of freedom
to these spin devices: by simply {changing} the top gates as low, high, and medium values,
spin devices for spin up only, %(with low top gates),
spin down only, %(with high top gates),
or both spins (occurring sequentially) %(with medial top gates)
can be respectively realized.
On the other hand, the gate-induced {region} tunability in the proposed device further adds
an output degree of freedom to these spin devices.
This may be especially important in {spin switchinges and spin memories},
whose lowest output voltage and two stored states
are found to be exactly determined by the position of the NDR {regions}
(see Ref. \cite{mizuta2006physics} and relevant references therein).

\subsection{Electrode doping effect}

%In all the above calculations, the effect of the electrode doping's ($U$) is ignored.

{
We at last investigate the electrode doping effect on the spin-dependent NDR.
%ligth electron and hole doping's are considered.
Fig. \ref{fig:electrode}(a,b) show the spin-dependent I-V characteristics at various light contact doping's. The spin-dependent PVR and NDR region ($eV_P^\pm,eV_V^\pm$) are summarized
in Fig. \ref{fig:electrode}(c,d) as a function of the contact doping.
A dual-spin NDR ($E_F=5.5$ and $V_t^{(i)}=2$) is considered.
As can be seen, the dual-spin NDR only maintains for -23meV$<U<$44meV.
For $U$ smaller than -23meV or larger than 44meV, the NDR
becomes spin-down only or spin-up only.
The single-spin NDR feature even disappears for $U$ smaller than -37meV or larger than 50meV.
%Actually, the ndr feature disappear when a doping smaller than -20 or larger than 45
%(smaller than -35 or larger than 40) is induced.
However, for familiar metal electrodes such as %Pd,
Ag, Ti, Cu, Au, and Pt at their \emph{equilibrium distances} to graphene,
the contact doping's are
-320meV, -230meV, -170meV, 250meV, and 320meV,
respectively.\cite{lee2008contact,giovannetti2008doping}
These values are much larger than the doping range ($-23$meV$<U<$44meV) the dual-spin NDR can survive.
%the spin-dependent Dirac gap in the ferromagnetic graphene.
%($\Delta_{+(-)}=13.4 (9.8) E_0$) \cite{song2017electric}.
%for familar metals at their ... distances with graphene, the doping's are
%... , which are rather ... to ... .
To maintain the proposed NDR, one possible way is to
%hence should
tune the metal-graphene distances to special values (e.g., Au/Cu/Ag at 3.2/3.4/3.7 $\AA$) to obtain ideal doping's.\cite{giovannetti2008doping}
%or a little bigger or a light hole doping's.
Another possible way is to adopt improved contacts such as Ti/Au. \cite{lee2008contact,robinson2011contacting}
}

{
%bias where the current peak and valley happen
In Fig. \ref{fig:electrode}(a,b) it is also clear that, both the peak and valley currents decrease with an increasing doping.
This is a result of that, the TMNs for the dominating resonant states
%in Eq. (\ref{eq:current})
($|E_\pm^+-U|=E_\pm^+-U>0$) decrease with the doping $U$.
However, the peak and valley currents decrease differently.
As shown in Fig. \ref{fig:dc}, the peak current is dominated by the resonant peak
%differential conductance
away from the Dirac point,
%over the whole integrating range;
%that totally insert the integrated window;
so as the Dirac point is lifted up it decreases
%first gentle and then shaply.
first gently and then sharply, see Fig. \ref{fig:electrode}(a);
%(guided by the dashe line in ... ).
in contrast, the valley current is dominated by the
resonant peak
%differential conductance
around the Dirac point, so as the Dirac point is lifted up it decreases
%first sharply and then gently
first sharply and then gently.
As a result of this contrast, the PVR %which is the most important parameter for ndr,
first increases and then decreases with an increasing doping,
see Fig. \ref{fig:electrode}(c).
(The spin-down PVR shows an exception for -10meV$<U<$0meV because
the resonant peak for the peak current extends considerably to the Dirac point.)
%becomes bigger when the doping is increaseda light hole doping is induced.
Importantly, the PVR for the spin-up NDR increases from 1.92 at an ideal doping
to 2.89 at a light hole doping of 22meV,
with a large enhancement factor of 51$\%$.
%show non- dependence on the doping.
%for spin up, the pvr becomes small when a electron doping is induced;
%the heavier the doping, the smaller the pvr.
%it becomes bigger when a small hole dopong is induced,
%smaller again for too big hole doping.
For spin down, the PVR of 1.15 at an ideal contacting approaches its maximum
of 1.27 at a light hole doping of 15meV, with an enhancement factor of 10$\%$.
The above specific metal-graphene distances can be enlarged a little
to achieve these light hole doping's\cite{giovannetti2008doping} and larger PVR's.
%bigger pvr can be obtained for electron doping larger than ...
%and hole doping smaller than ... .
%the maximum pvr for certain dopings the valley current decreases harder.
%For spin up (spin down) The PVR gets smaller when an electron doping ()
%or a hole dopoing larger than ... meV (...meV) are induced.
}

%%%
{
In Fig. \ref{fig:electrode}(d) it is seen that, the NDR regions are rather sensitive to
the contact doping's,
with variable ranges of 84 meV and 110 meV for spin up and spin down, respectively.
%For spin up (spin down), the central of the NDR window changes within a range of ... ().
The larger the contact doping, the lower the NDR regions.
This is because as $U$ increases, %the valley current arises as . as
the Dirac node in the source gets closer to the resonant levels
and the biases needed to %the resonant level
align them
%approach the Dirac node in the source
%or totally insert into the window
become smaller.
However, using the contact doping instead of the top gates to tune the NDR regions
would be much harder.}
%smaller biaes is needed.
%this is consistance with the mechanism the ndr arises.
%Hence it is important to investigate how the spin-dependent
%NDR can be affected by the contact doping.
%In Fig. \ref{fig:electrode}(a) we plot the calculated spin up I-V characteristic.
%It is seen that, the current decreases as the contact becomes non-ideal;
%the heavier the doping, the stronger the decrease.
%In surprise, the PVR can be enhanced for some positive contact doping, e.g., Cu,
%see Fig. \ref{fig:electrode}(b), where the $I_P$, $I_V$, and PVR for spin up are plotted.
%The PVR is enhanced from 1.92 at ideal contact to 2.33 at Cu contact,
%with an enhancement factor of 21$\%$.
%The case is similar for spin down, for which the PVR is plotted in Fig. \ref{fig:electrode}(b).
%%while for spin down
%The change of PVR due to contact is from 1.14 to 1.45,
%with an enhancement factor of 27$\%$.
%From Fig. \ref{fig:electrode}(b), it is also seen that,
%other metals with positive contact doping smaller than $\sim$240 meV
%can also be used for a larger PVR than the ideal contact case.

\section{Conclusion}\label{sec:IV}

In summary, we have proposed a bulk graphene based,
spin-dependent NDR device, which is composed of a sufficiently wide and short graphene and two
gated EuO strips deposited on top of it.
{An ideal or light hole electrode doping is also essential.}
The spin-dependent NDR arises because
%the biased transport is dominated by, whose
the energies (transversal mode numbers) of the spin-dependent resonant states decrease with an increasing bias.
Compared with spin-dependent NDR devices based on zigzag or armchair GNRs,
the proposed device holds the advantages of no need of edge tailors, %fewer regions,
generation of larger currents,
and most importantly, %the outstanding features of
features of a spin selectivity and a {region} tunability. %induced by top gates.
These remarkable features stem from a top-gate control of
% hence
the deviations of the spin-dependent resonant levels from the Fermi energy
and {the} Dirac point in the source electrode, respectively.
%The last two remarkable features
They add a spin and a bias degree of freedom to {conventional} NDR-based devices,
which paves a way for design of a whole new class of NDR circuits such as
spin-selectable and output-tunable spin switching and spin memory.
%%
%The proposed device can also be applied as a spin current filter at high bias.
%%%%%%
%Our work differs from similar proposals in that we use an ab initio
%electronic structure of ferromagnetic graphene and consider a biased transport.

\section*{Acknowledgements}

This work was supported by the National Natural Science Foundation of
China (NSFC) under Grant No. 11404300
%the Science Challenge Project (SCP) under Grant No. TZ2016003-1,
and the S$\&$T Innovation Fund of IEE, CAEP under Grant No. S20140807.

%%REFERENCES%%%
%\bibliography{sNDR} %You need to replace "rsc" on this line with the name of your .bib file
%\bibliographystyle{apsrev4-1} %the RSC's .bst file


\end{document}